\documentclass{aa}  
\usepackage{graphicx}
\usepackage{txfonts}
\def\o3            {{[O~{\small III}]}\/}

\def\Oc            {{\o3 \/$\lambda\/5007$}\/}
\def\hb            {{H$\beta$}\/}
\def\ha            {{H$\alpha$}\/}

\def\n2            {{[N~{\small II}]}\/}
\def\Na            {{\n2 \/$\lambda\/6548$}\/}
\def\Nb            {{\n2 \/$\lambda\/6583$}\/}

\def\flunits       {{ergs cm$^{-2}$ s$^{-1}$}\/}
\def\intunits      {{\flunits \,arcsec$^{-2}$}\/}

\begin{document}
\title{The luminous infrared composite Seyfert 2 galaxy NGC~7679 through the [O\,III] $\lambda\,5007$ emission line
         \thanks{
       Based on observations obtained at the Peak Terskol Observatory, Caucasus, Russia.}
       }
\author{I. M. Yankulova\inst{1}
        \and 
        V. K. Golev\inst{1}
        \and
        K. Jockers\inst{2}
        }
\offprints{Ivanka Yankulova, \email{yan@phys.uni-sofia.bg}} 
\institute{Department of Astronomy, St Kliment Okhridski University of Sofia, 
           5 James Bourchier Street, BG--1164 Sofia, Bulgaria\\
          \email{yan@phys.uni-sofia.bg, valgol@phys.uni-sofia.bg}
          \and
          Max-Planck-Institut f\"ur Sonnensystemforschung, Max-Planck-Stra{\ss}e 2,
          D--37191 Katlenburg-Lindau\\
          \email{jockers@linmpi.mpg.de}
          } 
\date{Accepted March 29, 2007}

 
  \abstract
  {  
NGC~7679 (Mrk 534) is a nearby ($z = 0.0177$) nearly face-on SB0 luminous infrared Sy2 galaxy in which starburst and AGN activities co-exist. The ionization structure is maintained by both the AGN power-law continuum and starburst. The galaxy is a bright X-ray source possessing a low X-ray column density $N_\mathrm{H} < 4\times10^{20}$ cm$^{-2}$. 
  }
  {
The Compton-thin nature of such unabsorbed objects infers that the simple formulation of the Unified model for SyGs is not applicable in their case. The absorption is likely to originate at larger scales instead of the pc-scale molecular torus. The main goal of this article is to investigate both gas distribution and ionization structure in the circumnuclear region of NGC~7679 in search for the presence of a hidden Sy1-type nucleus, using the \Oc \ luminosity as a tracer of AGN activity. 
  }
  {    
NGC~7679 was observed with the 2m RCC reflector of the Ukraine National Astronomical Observatory at peak Terskol, Caucasus, Russia. The observations were carried out on October 1996 with the Focal Reducer of the Max-Planck-Institut f\"ur Sonnensystemforschung, Germany. All observations were taken with tunable Fabry-Perot narrow-band imaging with spectral FWHM of the Airy profile $\delta\lambda$ between  3 and 4 \AA\ depending of the used wavelength. 
  }
  {  
The \Oc \ emission-line image of the circumnuclear region of NGC~7679 shows elliptical isophotes extended along the PA$\approx 80^\circ$ in the direction to the counterpart galaxy NGC~7682. There is a maximum of this emission which is shifted $\sim 4$ arcsec from the center defined by the continuum emission. The maximum of ionization by the AGN power-law continuum traced by \Oc /\ha \ ratio is displaced by $\sim\,13$ arcsec eastward from the nucleus.
The direction where high ionization is observed at PA$\approx 80^\circ \pm 10^\circ$ coincides with the direction to the companion galaxy NGC~7682 (PA$\approx 72^\circ$). On the contrary, at PA$\sim 0^\circ$ the ionization in the circumnuclear region is entirely due to hot stars.
  }
  {
Both the ratio $(N_\mathrm{ph}/N_\mathrm{ion})_{h\nu\,>\,55\,\mathrm{eV}} \approx 0.2 - 20 $ of the number $N_\mathrm{ph}$ of photons traced by \o3 \ to the number $N_\mathrm{ion}$ of high-energy ionizing photons and the presence of weak and elusive \ha \ broad wings indicate a hidden AGN. We conclude that the dust and gas in the high ionization direction PA\,$\approx 80^\circ$ has a direct view to the central AGN engine. This possibly results in dust/star-formation decay. A large fraction of the unabsorbed Compton-thin Sy2s with \o3 \ luminosity $\ga 10^{41}$ erg s$^{-1}$ possesses a hidden AGN source. 
  }


\authorrunning{Yankulova I., Golev V., and Jockers, K.}

\titlerunning{The luminous infrared composite Seyfert 2 galaxy NGC 7679 through ...}

\maketitle


\section{Introduction}

Luminous infrared galaxies (LIGs) are characterized by extreme IR luminosities $L_\mathrm{IR} \ga 10^{11} L_\odot$ at mid- to far-infrared (FIR) wavelengths. In their comprehensive spectroscopic survey of LIGs Kim et al. (\cite{Kim+95}) and Veilleux et al. (\cite{Ve+95}) have shown a clear tendency for the more luminous objects to be more Seyfert-like. The starburst and AGN are tightly connected phenomena and the interaction between them is a matter of debate.

Based on a large spectroscopic optical survey of bright IRAS and X-ray sources from ROSAT All Sky Survey, Moran et al. (\cite{Mo+96}) extracted low-redshift galaxies with optical spectra characterized by the HII regions and X-ray luminosities typical of AGNs and these objects were named Composite Seyfert/Starburst galaxies. Other similar galaxies (i.e. with bright X-ray emission together with the clear predominance of a starburst in the optical and IR regime) have been found also in the deep ROSAT fields (Boyle et al. \cite{Bo95}, Griffiths et al. \cite{Gr+96}) and in the Chandra and XMM-Newton deep fields (Rosati et al. \cite{Ros+01}). 

A significant part of the observed FIR-emission of these composites could be associated with circumnuclear starburst events. The nuclear X-ray source there is generally absorbed with column density of $N_\mathrm{H} > 10^{22}$ cm$^{-2}$ and these values range from $10^{22}$ cm$^{-2}$ to higher than $10^{24}$ cm$^{-2}$ for about 96 \% of this class of objects (Risaliti et al. \cite{Ri+99}, Bassani et al. \cite{Ba+99}). The circumnuclear\, starburst should also play a major role in the obscuration processes -- see for details Levenson et al. (\cite{Le+01}) and references therein. However, there are Sy2 galaxies with column densities lower than $10^{22}$ cm$^{-2}$. Panessa \& Bassani (\cite{PB02}, hereafter PB02) present a sample of 17 type 2 SyGs showing such low absorption in X-rays. The Compton thin nature of these sources is strongly suggested by some isotropic indicators such as FIR and \o3 \, emission. 

The fraction of Composite Seyfert/Starburst objects is estimated to be in the range of 10$\%$ - 30$\%$ of the Sy2 population. The simple formulation of the Unified model for SyGs is not applicable in such sources. The observed absorption is likely to originate at larger scales instead in the pc-scale molecular torus. Probably the Broad Line Regions (BLRs) of these objects are covered by some obscuring dusty material.

NGC~7679 is a nearby ($z = 0.0177$) nearly face-on SB0 Seyfert 2 type galaxy in which starburst and AGN activities co-exist. The IRAS fluxes show that the luminosity of NGC~7679 in the far infrared is about $L_\mathrm{FIR} \approx 10^{11} L_\odot$. This object is included in the large spectroscopic survey of 200 luminous IRAS galaxies (Kim et al. \cite{Kim+95}, Veilleux et al. \cite{Ve+95}). NGC~7679 is physically associated by a common stream of ionized gas with the Sy2 galaxy NGC~7682 at $\sim\,4.5$ arcmin eastward (PA $\approx$ 72$^\circ$) forming the pair Arp 216 (VV 329). The tidal interactions between both galaxies together with the existence of a bar in NGC~7679 could enhance the gas flow towards the nuclear regions and possibly trigger the starburst processes (Gu et al. \cite{Gu+01}).

The X-ray properties of the NGC~7679 based on the BeppoSAX observations and on the ASCA archive were discussed by Della Ceca et al.  (\cite{Ce+01}, hereafter DC01). Their conclusion is that NGC~7679 is a Seyfert-starburst composite galaxy which implies the clear predominance of an AGN in the X-ray regime connected with a starburst in the optical and IR regime. DC01 found that a simple power-law spectral model with $\Gamma \sim 1.75$ and small intrinsic absorption ($N_\mathrm{H} < 4 \times 10^{20}$ cm$^{-2}$) provides a good description of the spectral properties of NGC~7679 from 0.1 to 50 keV. The small X-ray absorption and the absence of strong (EW $\sim$ 1 keV) Fe-lines suggest a Compton thin type 2 AGN in NGC~7679 which clearly distinguishes this galaxy from the other LIG Seyferts. 

The main goal of this article is to investigate both gas distribution and ionization structure in the circumnuclear regions of the luminous IR unabsorbed Seyfert galaxy NGC~7679 and to look for tracers of the presence of a hidden Sy1-type nucleus. 

Some information on the observations and data reduction procedures is presented in Section 2. The results are presented in Section 3 and discussed in Section 4. The combination of the data taken from recent literature and our Fabry-Perot observations provides new insight in the circumnuclear region of NGC~7679 and in the phenomena occurring there.

\section{Observations and data reduction} 

   \begin{table}
      \caption[]{Observiation details}
         \label{tab1}
     $$ 
         \begin{tabular}{llll}
            \hline
            \hline \\
            \noalign{\smallskip}
           image & interference                  & Fabry-Perot           & frames    \\
           frame & filter$^{\mathrm{a)}}$        & tuned                 & $\times$  \\ 
                 &                               & wavelength            & exposure  \\
                 & $\lambda_c/\mathrm{FWHM}$ & $\lambda_\mathrm{FP}$ & time      \\ [2.5mm]
                 & (\AA)/(\AA)                   & (\AA)                 & (s)       \\
                 &&\\ \hline \\ 
\ha                     &  6662/55   & 6674.8 & 1 $\times$ 1800 \\[1.5mm]
                        &            & & 2 $\times$ 900  \\[1.5mm]
\Na                     &  6662/55   & 6659.9 & 1 $\times$ 900  \\[1.5mm]
continuum               &  6719/33   & 6720.0 & 1 $\times$ 1800 \\[1.5mm]
                        &            & & 1 $\times$ 900  \\[1.5mm]
\Oc                     &  5094/44   & 5092.4 & 2 $\times$ 900  \\[1.5mm]
continuum               &  5002/41   & 4437.7 & 1 $\times$ 1200 \\[1.5mm]
Gunn r$^{\mathrm{b)}}$  &  6800/1110 & & 1 $\times$ 60   \\[1.5mm]
BG 39/2$^{\mathrm{b)}}$ &  4720/700  & & 2 $\times$ 1500 \\[1.5mm]
            \hline
            \hline            
            \noalign{\smallskip}
         \end{tabular}
     $$ 
\begin{list}{}{}
\item[$^{\mathrm{a)}}$] Used to separate Fabry-Perot working orders
\item[$^{\mathrm{b)}}$] Broad-band image taken without Fabry-Perot to reveal the morphology
\end{list}
   \end{table}

NGC~7679 was observed by K. Jockers, T. Bonev, and T. Credner with the 2m RCC reflector of the Peak Terskol Observatory, Caucasus, Russia. The observations were carried out in October 1996 with the Two-channel Focal Reducer of the former Max-Planck-Institut f\"ur Aeronomie, Germany (now Max-Planck-Institut f\"ur Sonnensystemforschung, MPS). This instrument was primarily intended for cometary studies but it has repeatedly been used for observations of active galactic nuclei (see for example Golev et al. \cite{G+95}, \cite{G+96}, and Yankulova \cite{Yan99}). The technical data and the present capabilities of the MPS Two-channel Focal Reducer are described in Jockers (\cite{Jo97}) and Jockers et al. (\cite{Jo+00}). 

All observations were taken in Fabry-Perot (FP) mode using tunable FP narrow-band imaging with spectral FWHM of the Airy profile $\delta\lambda$ in order of  3 - 4 \AA . The details of observations are presented in Table~\ref{tab1} where the central wavelengths $\lambda_c$ and the effective width $\Delta\lambda$ of the interference filters used to separate the Fabry-Perot interference orders, the wavelength $\lambda_\mathrm{FP}$ at which the Fabry-Perot was tuned, and the exposures are listed.

The overall ``finesse'' of the system $\Delta\lambda/\delta\lambda$ is $\approx$ 15, $\Delta\lambda$ is the free spectral range of the FP. As one can see from Table~\ref{tab1} $\Delta\lambda$ is comparable to the filter's band width and therefore all FP orders except the central one are efficiently suppressed. Two exposures of NGC~7679 were obtained through each filter to eliminate cosmic ray events and to increase the signal-to-noise ratio. 
Flat­field exposures were obtained using dusk and dawn twilight for uniform illumination of the detector. No dark correction was required.

The images were reduced following the usual reduction steps for narrow-band imaging. After flat­fielding the frames were aligned by rebinning to a common origin. The final alignment of all the images was estimated to be better than 0.1 px (the scale is 1 px = 0.8 arcsec). A convolution procedure was performed in order to match the Point-Spread Functions (PSFs) of each line-continuum pair which unavoidably degrades the final FWHM of the images to the mean value $\approx 3 - 3.3$ arcsec (shown as 'seeing' in Fig.~\ref{fig1}). At the distance of NGC~7679 one arcsec corresponds to a distance of about 340 pc assuming $H_0 = 75$ km sec$^{-1}$ Mpc$^{-1}$.


\section{Results} 

\subsection{Narrow-band emission-line images} 


   \begin{figure} 
      \centering
      \includegraphics[width=9cm]{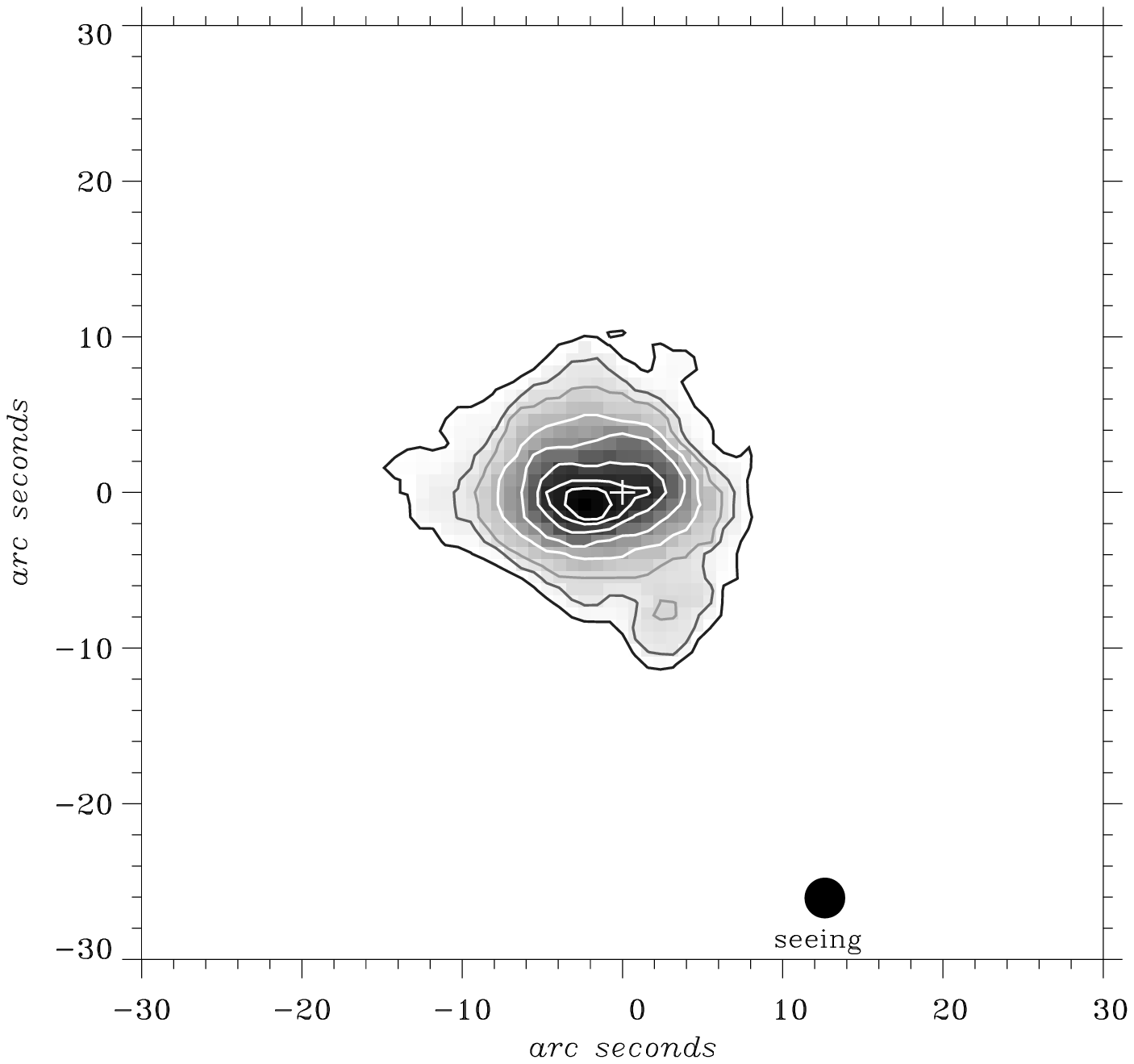}
      \caption{ 
                Contours of continuum-subtracted narrow-band \Oc -image 
                superimposed on the gray-scale \Oc -emission distribution of the  
                circumnuclear region of NGC~7679. 
                The background noise level is $\sigma = 2.01\times10^{-17}$ \intunits .
                The outermost contour is taken at 5$\sigma$ above the sky level and 
                the next contours increase by a factor of $\sqrt{2}$. 
                Note East-West elongation and two extrema decentered of about $\sim\,4$ arcsec 
                from the position of the nucleus marked by cross. North is up, East is to the left.
              }
      \label{fig1}
   \end{figure}
%


   \begin{figure} 
      \centering
      \includegraphics[width=9cm]{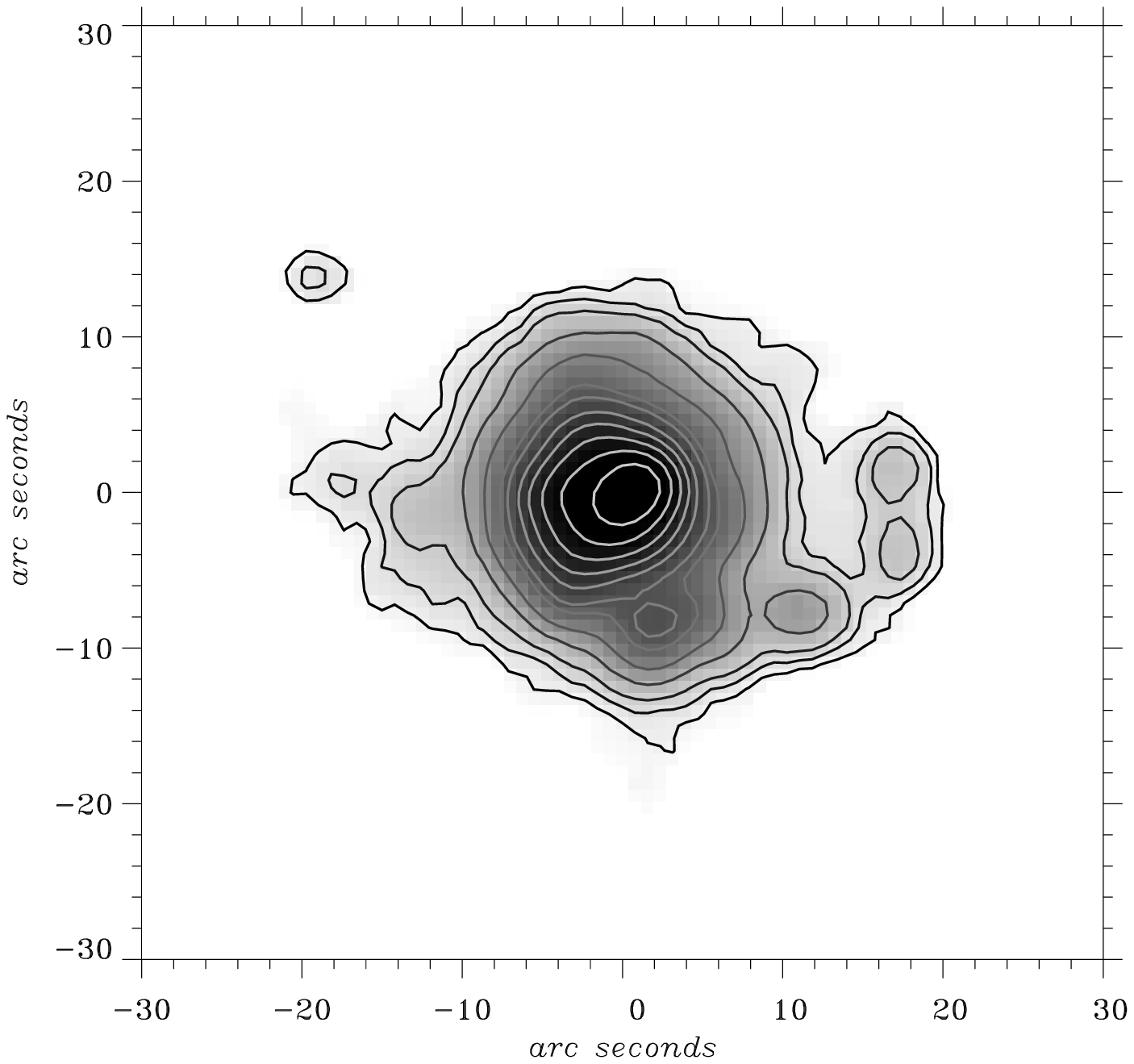}
      \caption{ Contours of continuum-subtracted narrow-band \ha -image 
                superimposed on the gray-scale \ha -emission distribution of the  
                circumnuclear region of NGC~7679.
                The background noise level is $\sigma = 2.77\times10^{-18}$ \intunits .
                The outermost contour is taken at 5$\sigma$ above the sky level and 
                the next contours increase by a factor of $\sqrt{2}$. North is up, East is to the left.
              }
      \label{fig2}
   \end{figure}
%


   \begin{figure} 
      \centering
      \includegraphics[width=9cm]{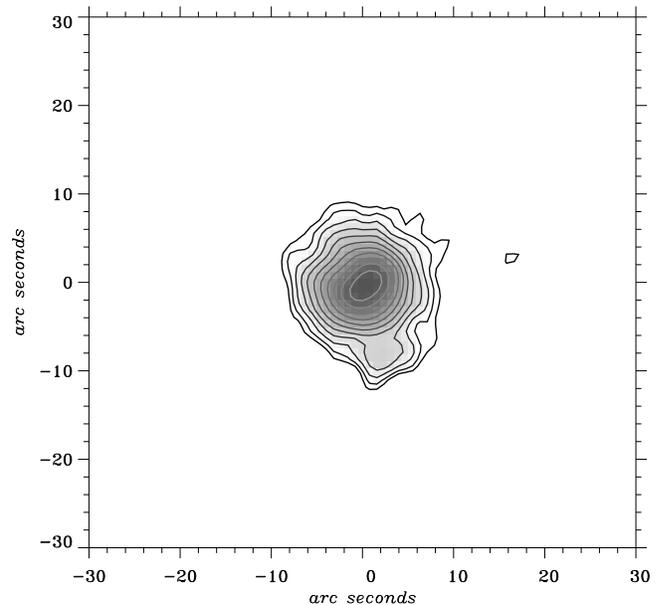}
      \caption{Contours of continuum-subtracted narrow-band \Na -image 
               superimposed on the gray-scale \Na -emission distribution of the  
               circumnuclear region of NGC~7679. 
               The background noise level is $\sigma = 4.75\times10^{-18}$ \intunits .
               The outermost contour is taken at 5$\sigma$ above the sky level and 
               the next contours increase by a factor of $\sqrt{2}$. North is up, East is to the left.
              }
      \label{fig3}
   \end{figure}
%

Gray-scale images of the narrow-band flux distribution of the extended circumnuclear region of NGC~7679 in the \Oc, \ha, and \Na \ emission lines with superimposed contours are presented in Fig.~\ref{fig1}, \ref{fig2}, and \ref{fig3}, respectively. 

The \Oc\ emission shown in Fig.~\ref{fig1} reveals a bright, about 20 arcsec in size, extended emission-line region (EELR) which is elongated approximately in East direction (PA $\approx\, 80^\circ \pm 10^\circ$).  This region is similar to the analogous EELRs observed in many Sy2 type galaxies. Most probably it is powered by the AGN-type activity of the nucleus. The emission-line peak of \Oc \, is shifted at about $\sim 4$ arcsec to the East with respect to the center defined by the continuum emission and marked by cross in Fig.~\ref{fig1}.

At larger distances ($\sim $ 37 arcsec) the ionized gas forms an envelope which is extended along the direction PA $\approx$ 72$^\circ$ to the NGC~7682, the counterpart of NGC~7679, as it was already noted by Durret \& Warin (\cite{DW90}).

In Fig.~\ref{fig2} we present our very deep and high-contrast \ha\ continuum-subtracted image with numerous starburst regions where because of both seeing and pixel size we are able to see only elliptical central isophotes instead of the ``double nucleus'' observed recently by Buson et al (\cite{Bu+06}). Our analysis of the unpublished \ha\ images taken from the archive of the Isaak Newton Group of telescopes at La Palma as well as the archive images of Buson et al. (\cite{Bu+06}) from the ESO La Silla NTT also revealed a ``double nucleus'' otherwise unseen in the known broad-band images. The separation between the nuclear counterparts (in fact one is the active nucleus itself and the other one is a bright spiral-like extremely powerful starburst region) is $\approx 3$ arcsec. The existence of this ``double nucleus'' in NGC~7679 could enhance the gas flows towards the nuclear regions and possibly trigger the starburst process itself. The ``double nucleus'' can be also seen at very different wavelength range on 6 cm and 20 cm high-resolution VLA radio continuum map of NGC 7679 published by Stine (\cite{St92}). The angular distance and PA between two counterparts is quite the same. The radio spectral index is $-0.37$ and steepens away from the center which indicates that nonthermal emission leaks out of the starburst region.

The low-excitation gas traced by the emission in \ha \ reveals different morphology as compared to that of the \Oc \ emission. Inside of the region with radius of 6 -- 8 arcsec from the center the contours of the \ha\ emission are nearly circular. Outside this region to the West of the main body of NGC~7679 a clearly outlined wide arc is observed at 16 arcsec ($\sim$ 5 kpc) from the center. To the East this arc converts into a gaseous envelope which forms a part of a circumnuclear starforming ring mentioned by Pogge (\cite{Po89}). This arc is not detected on the narrow-band continuum image next to the \ha. The same morphology in \ha \ + \n2 \ with higher spatial resolution was observed by Buson et al. (\cite{Bu+06}).

The Fabry-Perot technique used by us makes possible to disentangle \Na\ from \ha. The pure \Na\ emission (Fig.~\ref{fig3}) shows extended structure $\sim\,20$ arcsec in diameter. The starforming ring revealed by the \ha\ image is not seen here. As a rule the gas component in the starforming ring is ionized by stellar UV-emission and the \Na \ is weaker than that one where the gas is ionized by power-low AGN continuum.
On the other hand, this could be an effect due to the shorter exposure time of our \Na \ frame.

\subsection{Narrow-band emission-line total fluxes}  

The total emission-line fluxes of \ha , \Oc  and \Nb \ were estimated from our flux calibrated images in an aperture of 2 kpc ($r\,\la\,3$ arcsec) like the one used by the authors cited in Table~\ref{tab2}. In this Table we have collected available measurements of the emission lines observed by us up to now. Our measured fluxes are in good agreement with those of Kim et al. (\cite{Kim+95}) and differ from the measurements of Contini et al. (\cite{Cont+98}). Flux values given by Contini et al. (\cite{Cont+98}) are twice larger than ours and those given by Kim et al. (\cite{Kim+95}). 

Recently Gu et al. (\cite{Gu+06}) measured the central flux in \Oc . We found a reasonable  coincidence between their value ($1.55 \times 10^{-14}$ \flunits ) and ours ($1.94 \times 10^{-14}$ \flunits ) in the much smaller aperture used by them.

   \begin{table*}
      \caption[]{Measured emission lines fluxes in 2 kpc central aperture in NGC~7679                   }
         \label{tab2}
     $$ 
         \begin{tabular}{llllllll}
            \hline
            \hline \\
            \noalign{\smallskip}
Emission & \multicolumn{7}{c}{Measured flux $F$($\lambda$), ergs cm$^{-2}$ s$^{-1}$} \\[2mm]
         & \multicolumn{5}{c}{ 2 kpc central aperture} &
           \multicolumn{2}{c}{9 arcsec off the nucleus}\\[2mm]
         & 1 & 2 & 3 & 4 & 5 & 6 & 7 \\ 
\hline\\ 

H$\alpha$ & $1.92\times10^{-13}$ & $1.9\times10^{-13}$  & $4.5\times10^{-13}$ & -- & $3.8\times10^{-13}$ &
            $3.73\times10^{-14}$ & $1.04\times10^{-14}$ \\[6pt]
\Na       & $9.96\times10^{-14}$ & $1.08\times10^{-13}$ & $1.86\times10^{-13}$ & -- & -- &
            $9.8\times10^{-15}$ & $4.5\times10^{-15}$ \\[6pt]
\Oc       & $5.2\times10^{-14}$ & $5.3\times10^{-14}$ & $8.8\times10^{-14}$ & -- & -- &
            $9\times10^{-15}$ & $4.6\times10^{-15}$ \\[6pt]
H$\beta$  & -- & $1.1\times10^{-14}$ & $5.24\times10^{-14}$ & $1.0\times10^{-14}$ & -- &
            $5.9\times10^{-15}$ & -- \\[12pt]  
$F$(H$\alpha$)/$F$(H$\beta$) & -- & 17.4 & 8.5  & 5.0  & 4.58 & 6.3 & -- \\[6pt]
$F$(H$\gamma$)/$F$(H$\beta$) & -- & 0.24 & 0.32 & 0.4  & 0.3 & --  & -- \\[12pt] 
$C$        & -- & 4.93  & 2.88 & 1.6  & 1.12 & 2.02 & -- \\[6pt]
$E(B-V)$ & -- & 1.45  & 0.85 & 0.47 & 0.33 & 0.65 & -- \\[12pt]
            \hline
            \hline            
            \noalign{\smallskip}
         \end{tabular}
     $$ 
\begin{list}{}{}
  \item[] Columns: 1 - this work; 2 - Kim et al. (\cite{Kim+95}); 3 - Contini et al. (\cite{Cont+98});
                   4 - Kewley et al. (\cite{Ke+00}); 5 - Buson et al. (\cite{Bu+06}); 
                   6 - Contini et al. (\cite{Cont+98}); 7 - this work, PA = $207^\circ$.
\end{list}
   \end{table*}

We estimated the flux of the continuum near \Oc\ within the central 2 kpc to be
$F (\lambda_\mathrm{cont})=6.74\times10^{-15}$\,\flunits \AA . Then the equivalent width of the emission line \Oc \ is $EW(\lambda\,$5007) = 7.6 \AA . Baskin and Loar (\cite{BL05}) have used the photoionization code CLOUDY to calculate the dependence of $EW(\lambda\,$5007) on the electron density $n_\mathrm{e}$, the ionization parameter $U$, and the covering factor $CF$. 
Following their Fig. 5 and our estimation of $EW(\lambda\,5007)$ we derive for the covering factor $CF$ the range $0.016 \le CF \le 0.04$ with the most probable value $CF \approx 0.024$. 
 
There is a large quantity of absorbing matter in the central region of NGC~7679 (Telesco et al. \cite{T+93}) which modifies the Balmer emission lines. The Balmer decrement reported by Kim et al. (\cite{Kim+95}) in the central 2 kpc is $F$(\ha )/$F$(\hb )$\,\approx\,$17.4, but following Contini et al. (\cite{Cont+98}) this decrement is 8.5. Kewley et al. (\cite{Ke+00}) give $E(B-V)$\,=\,0.47 which results to $F$(\ha )/$F$(\hb )$\,=\,$5.04. In Table \ref{tab2} the value of the parameter $C$ is evaluated from the measured Balmer decrement and from the assumption that in AGNs $F$(\ha )/$F$(\hb )$\,=\,$3.1 and the optical depth $\tau_{\lambda} = C f(\lambda)$ where $f(\lambda)$ is the reddening curve (Osterbrock \cite{Os89}). The extinction $E(B-V)$ derived from the Balmer decrement is also given in Table~\ref{tab2}.

Contini et al. (\cite{Cont+98}) present measurements of emission-lines fluxes made in the extranuclear region 9 arcsec off the nucleus at PA $= \,207^\circ$ in an aperture of 3 arcsec.
We estimated the emission-line fluxes from our images in the same aperture at the same place in order to compare with those given by Contini et al. (\cite{Cont+98}). The results are given in Table 2. The Contini's values are about 2 times larger than ours in the extranuclear region as well as at the nucleus.

Moustakas \& Kennicutt (\cite{MK06}) report total emission-line fluxes of \ha \ and \Oc \ in a wide rectangular aperture $30 \times 80$ arcsec oriented at PA = 90$^\circ$. Their \ha -flux $F$(H$\alpha$) = 
$(1.535 \pm 0.062) \times 10^{-12}$ \flunits \ coincides with our value ($1.52\times 10^{-12}$ \flunits ) in the same wide aperture after a correction for extinction with $E(B-V)$ = 0.065 used by them. In \Oc \ the coincidence is reasonably good ($4.72 \times 10^{-13}$ compared with ours $3.90 \times 10^{-13}$ \flunits ). 

\subsection{The ionization map $F$(\Oc )\,/\,$F$(\ha )}
   

   \begin{figure*} 
      \centering
      \includegraphics[width=8.5cm]{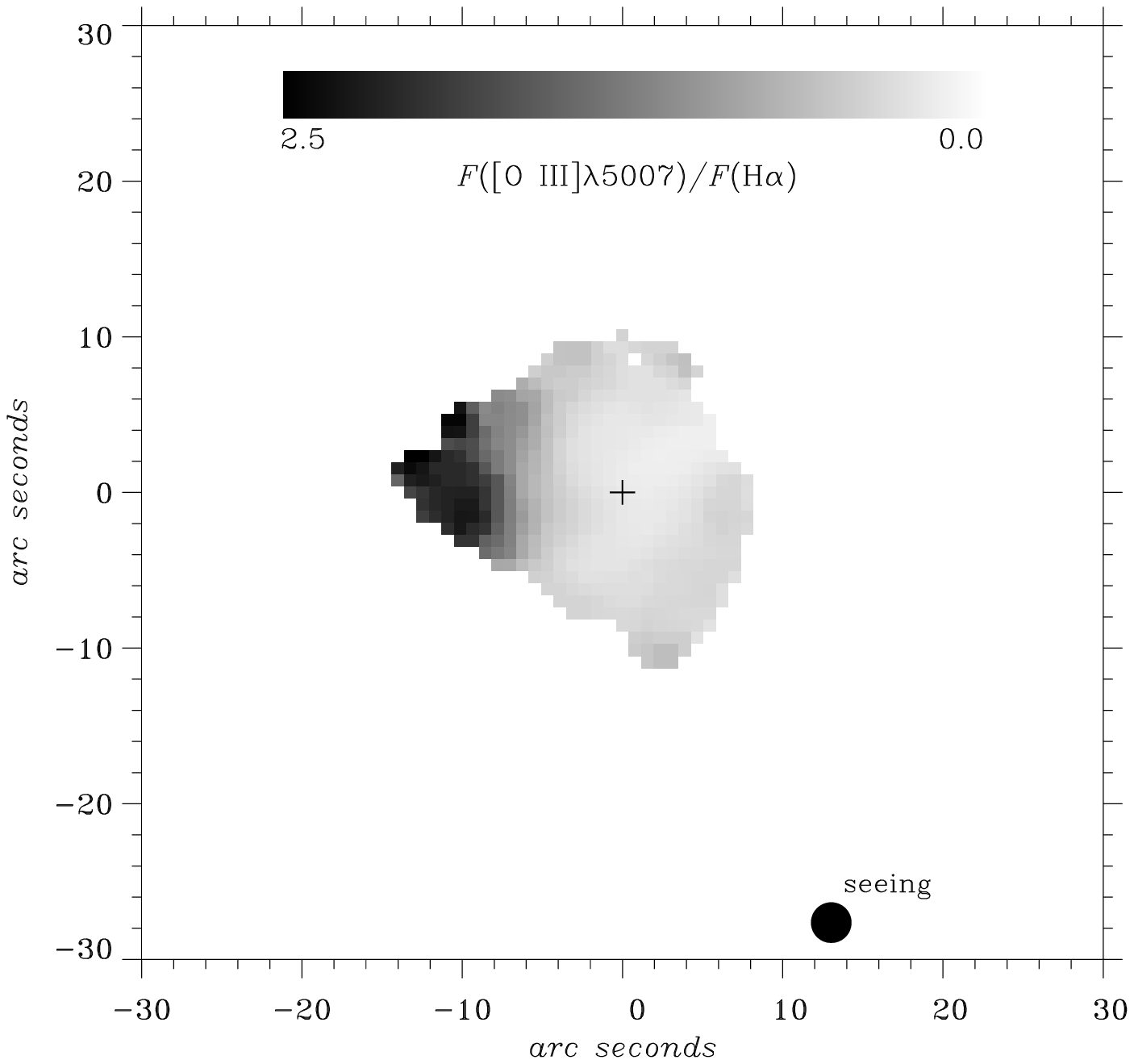}
      \includegraphics[width=8.5cm]{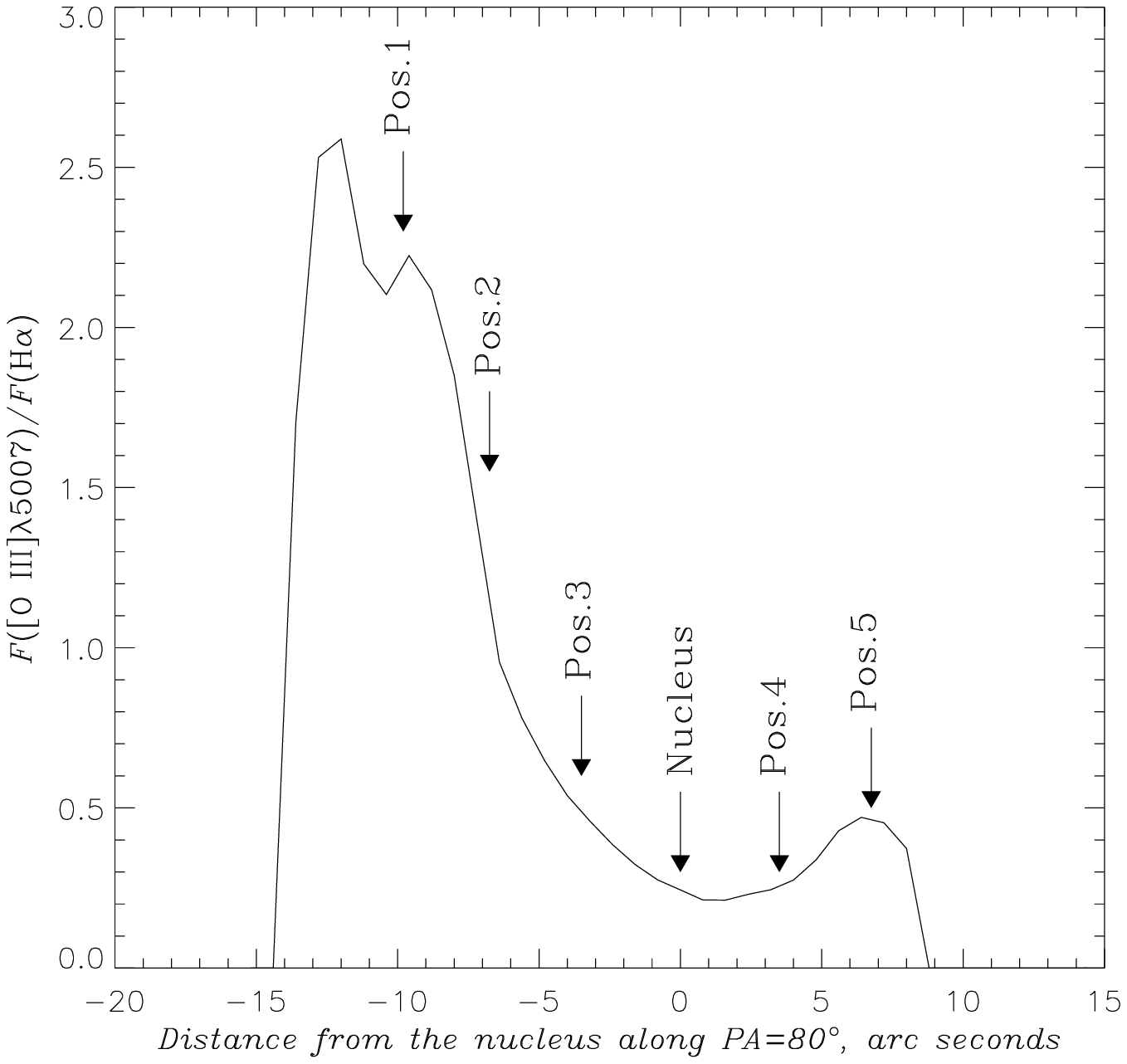}
      \caption{ $F$(\Oc )\,/\,$F$(\ha ) ionization map of NGC~7679. 
                All pixels below 4$\sigma $ of the background noise level were suppressed   
                before image division (left). The ratio $F$(\Oc )/$F$(\ha ) vs the axial
                distance from the nucleus along PA $\approx$ 80$^\circ$ (right). The 
                positions labeled 1 to 5 are equidistant with step size of 3 arcsec. 
                We refer to them later in the text (see Fig.~\ref{fig6}).
              }
      \label{fig4}
   \end{figure*}
%

Our flux-calibrated emission-line images are used to form the $F$(\Oc )/$F$(\ha ) ionization map in order to analyse the mean level of ionization. This map is shown in the left panel of Fig.~\ref{fig4}. All pixels below 4$\sigma$ of the background noise level were suppressed before the division of the corresponding images.  The ionization map infers a presence of a maximum shifted to the East at PA $\approx$\ 80$^\circ$ with respect to the photometric center defined by the integral light of the continuum images and marked by cross on the figure.

A slice of this map along the PA $\approx$\ 80$^\circ$ versus the axial distance from the nucleus is presented in the right panel of Fig.~\ref{fig4}. Below we will discuss in more detail the behaviour of the ionization at positions 1 to 5.

\section{Discussion} 

\subsection{The ionizing flux from the central engine}
 
In order to estimate the number of ionizing photons emitted from the central engine, we made use of the recent X-ray observations of NGC~7679. This object was observed by ASCA and BeppoSAX in 1998, and by XMM-Newton in 2005. A detailed analysis of ASCA and BeppoSAX data sets is present in DC01. They show that a single absorbed power-law function (with a photon index 1.75) fits the observed spectrum very well and the X-ray absorption is relatively small ($N_\mathrm{H}\leq 4\times10^{20}$~cm$^{-2}$). 

The data for the X-ray observations in 2005 were taken from the XMM-Newton public archive. The corresponding X-ray spectra for the PN and the two MOS detectors were extracted following the standard procedures using the XMM-Newton Science Analysis System software (SAS version 7.0.0). A single absorbed power-law function gave a good fit ($\chi^2/dof = 201/191$) to all the three spectra which were fitted simultaneously. The small X-ray absorption in the nucleus of NGC~7679 was confirmed, $N_\mathrm{H} = 5.6[4.0 \div 7.5]\times10^{20}$~cm$^{-2}$, and no
change in the shape of the spectrum was found, a photon index of $1.81[1.70 \div 1.92]$ (the 90\%-confidence intervals are given in brackets). The absorbing X-ray column density along the line of sight is about an order of magnitude smaller than that one estimated from the observed Balmer decrement which is $N_\mathrm{H}\sim 8 \times 10^{21}$ cm$^{-2}$ and $ \sim 5\times10^{21}$ cm$^{-2}$ following  Kim et al. (\cite{Kim+95}) and Contini et al. (\cite{{Cont+98}}), respectively.

Interestingly, the observed X-ray flux has decreased by a factor $\sim 10$ over a time period of $\sim 7$ years: $F_X = 3.8\times10^{-13}$~and~$5.8\times10^{-13}$~ergs cm$^{-2}$~s$^{-1}$ correspondingly in the 0.1-2.0 keV and 2.0-10.0 keV energy intervals. Since, on the one hand, there is only about 5\% scatter of the fluxes for all the three detectors
(one PN and two MOS) around the average values given above, and, on the other hand, NGC 7679 shows an appreciable X-ray variability (DC01), it is then likely that the detected decrease of the X-ray flux is real and not an instrumental effect. 

The extrapolation of the DC01's power law to the UV spectral domain (that is to $h\nu_0 = 13.6$~eV) yields
$F^\mathrm{nt}_\nu = F_{\nu_0} (\nu_0/\nu)^\alpha$ where $\alpha = 0.75$ and $F_{\nu_0}=2.0\times 10^{-28}$ erg cm$^{-2}$ s$^{-1}$ Hz$^{-1}$. 
The same extrapolation for the XMM-Newton spectrum results in $F^\mathrm{nt}_\nu = F_{\nu_0} (\nu_0/\nu)^\alpha$ where $\alpha = 0.81$ and $F_{\nu_0}=2.8\times 10^{-29}$ erg cm$^{-2}$ s$^{-1}$ Hz$^{-1}$. 

The number of ionizing photons with $h{\nu}>55$\,eV provided by the central AGN source is defined as  

 \begin{equation}
  N_\mathrm{ion} = \int\limits^\infty_{55\ \mathrm{eV}}  \frac{F\,^\mathrm{nt}_{\nu}}{h{\nu}} \, d\nu = 
  4{\pi}R^2_G\, \frac{F^\mathrm{nt}_{h\nu = 55\,\mathrm{eV}}} {h \alpha}
 \end{equation}
 
\noindent where  R$_\mathrm{G}$ is the distance to the NGC~7679. For the BeppoSAX data this estimation is $N_ \mathrm{ion}\sim 10^{52}$ ph s$^{-1}$ and for the XMM-Newton data $N_\mathrm{ion}\sim 10^{51}$ ph s$^{-1}$. These values are averaged between all BeppoSAX and XMM-Newton bands, respectively. The number of ionizing photons decrease from the BeppoSAX time to the XMM-Newton time in the range of $10^{51} \la N_\mathrm{ion} \la 10^{52}$ ph s$^{-1}$.

\subsection{Physical conditions in the circumnuclear region of NGC~7679}  

The extended emission-line region in NGC~7679 has a rather different morphology when observed in \ha\ (low ionization emission line) as compared to \Oc\ (high ionization emission line). The \ha \ image (Fig.~\ref{fig2}) contains a compact circumnuclear region ($\sim$ 20 arcsec in diameter) whose isophotes do not infer any preferred direction.
In contrast, the \Oc\ image (Fig.~\ref{fig1}) of the circumnuclear region of NGC~7679 shows elliptical isophotes extended along the PA $\approx\, 80^\circ \pm 10^\circ$. Such difference in morphology of the emission-line images signals the presence of at least two distinct ionization components (see for example Pogge \cite{Po89}). 

The extended morphology both of the \Oc \ image (Fig.~\ref{fig1}) and of the \Oc /\ha \ flux ratio image  (Fig.~\ref{fig4}) suggests an anisotropy of the radiation field. In order to check whether the ionizing field is collimated or not we have to compare the number of ionizing photons $N_\mathrm{ph}$, absorbed by the extended emission line  gas with the number of ionizing photons $N_\mathrm{ion}$, emitted by the central AGN engine. Usually, the hydrogen line flux $F$(\ha \,) or $F$(\hb \,) is used to find $N_\mathrm{ph}$. But the NGC~7679 high resolution \ha \ image reveals a central circumnuclear star-forming spiral ring capable of producing about $\sim$ 75\% of the optical line emission within a radius of $\sim1$\,kpc (Buson et al. \cite{Bu+06}). For this reason it is not quite correct to use the $F($\ha ) in order to make the $N_\mathrm{ph}$ estimate. 

Kauffmann et al. (\cite{Kau2003}) focus on the luminosity of the \Oc \ as a tracer of AGN activity. 
We can estimate the number  $N_\mathrm{ph}$ of ionizing photons with energy above $h\nu $ = 55\,eV from the observed  \Oc \ luminosity after correction for extinction. A dust correction to \o3 \ based on the ratio $F$(\ha )\,/\,$F$(\hb )\, should be regarded as best approximation (Kauffmann et al. \cite{Kau2003}). According to Draine \& Lee (\cite{DL+1984}) (Fig.7 therein) the optical depth is $\tau_{5007} = 0.96\,C = 2.76 $. Here we adopt the value of $C$= 2.88 following Contini et al. (\cite{Cont+98}) as a more compromising reddening value among the different Balmer decrement assessments. Then the luminosity, corrected for extinction, $L^\mathrm{corr}([\mathrm{O}^{+2}]\lambda5007) = 4.4 \times10^{41} $ {{ergs s$^{-1}$}. We note that PB02 give 5.7$\times10^{41}${{ergs s$^{-1}$}\, for the \Oc \, luminosity.

The total number of ionizing photons that must be available to produce the observed \Oc \, emission is given by the expression 

 \begin{displaymath}
     N_\mathrm{ph} = \frac{{\alpha}_G(\mathrm{O}^{+2},T_\mathrm{e})
                     L^\mathrm{corr}([\mathrm{O}^{+2}]\lambda5007)\,CF^{-1}}
                     {\alpha^\mathrm{eff}_\mathrm{5007}(n_\mathrm{e},T_\mathrm{e})\,   
                      h{\nu_{5007}}} 
 \end{displaymath}

 \begin {equation}            
     \ \approx 2\times10^{52}\ \mathrm{ph\,s}^{-1} \\
 \end{equation}
   
\noindent where $\alpha_G(\mathrm{O}^{+2},T_\mathrm{e})=5.1\times 10^{-12}$ cm$^3$ s$^{-1}$ (Aldrovandi \& Pequignot \cite{AP73}) is the recombination coefficient at $T_\mathrm{e}\approx10^4$ K and
$\alpha^\mathrm{eff}_\mathrm{5007}(n_\mathrm{e},T_\mathrm{e})= 1.1\times10^{-9}$ cm$^3$ s$^{-1}$ is the effective recombination coefficient at $n_\mathrm{e}=10^5$ cm$^{-3}$ and  $T_\mathrm{e}=10^4$ K. This coefficient strongly depends on the electron density and temperature. If we accept $T_\mathrm{e}=10^4$ K then 
$\alpha^\mathrm{eff}_\mathrm{5007}(n_\mathrm{e})= 5.14\times10^{-3} A_{21}/n_\mathrm{e}$\,
cm$^3$ s$^{-1}$ where A$_{21}$ =\,\,0.021 s$^{-1}$. As the critical electron density 
is $n_\mathrm{e}^\mathrm{cr}(5007) = 5\times 10^5$ cm$^{-3}$ we assume that the electron density is not lower than $n_\mathrm{e}\approx 10^4$ cm$^{-3}$ in order to emit the \Oc . Then the lower limit for $N_\mathrm{ph}$ is  $\approx 2\times10^{51}$ ph s$^{-1}$. For NGC~7679 the covering factor CF = 0.024. 

The photon ratio $N_\mathrm{ph}/N_\mathrm{ion}$ is a probe of the collimation hypothesis. In the anisotropic case this ratio is considerably larger than 1. Under the above assumptions about n$_\mathrm{e}$ and T$_\mathrm{e}$
we estimate for NGC~7679\, $0.2\la (N_\mathrm{ph}/N_\mathrm{ion})_{h\nu > 55\,\mathrm{eV}}\la 20$ but the lower limit could increase if the luminosity $L([\mathrm{O}^{+2}]\lambda5007)$ is integrated over the whole image.
The increase of the upper limit of this ratio is due to the XMM-Newton data which are $\sim $8 times lower than ASCA/BappoSAX ones. 
%

   \begin{figure} 
      \centering
      \includegraphics[width=9cm]{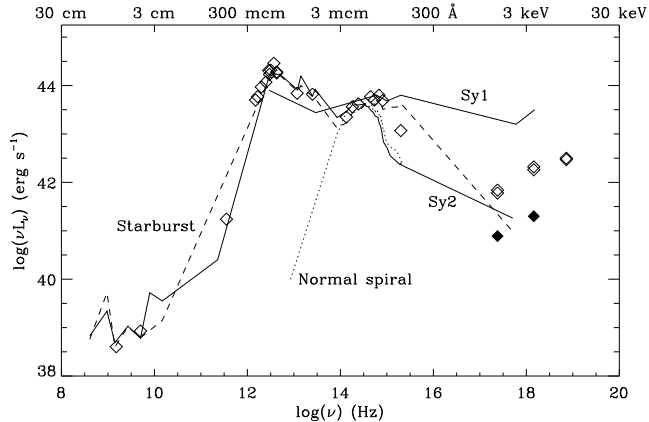}
      \caption {
                Spectral energy distribution (SED) from the radio to the X-ray band of the composite Starburst/Sy2 
                galaxy NGC~7679 (open diamonds). The radio values at 6 cm and 20 cm are from VLA (Stine, \cite{St92}). 
                The X-ray band data are from ASCA and BeppoSAX (DC02 and Risaliti, \cite{{Ri02}}). 
                Filled diamonds represent recent X-ray observations taken from the XMM-Newton archive.  
                All other data are taken from NED.
                The SED has been compared with a normal spiral galaxy template (dotted line) taken
                from Elvis et al.(\cite{E+94}), with Starburst and Sy2 galaxy templates (dashed line and thin solid
                line) taken from Schmitt et al. (\cite{Sch+97}), and with Sy1 galaxy template (thick solid line) 
                taken from Mas-Hesse et al. (\cite{M-H+95}).
               }
      \label{fig5}
   \end{figure}
%

Both the ratio $(N_\mathrm{ph}/N_\mathrm{ion})_{h\nu > 55\,\mathrm{eV}}$ and the presence of weak and elusive broad \ha -wings (Kewley et al. \cite{Ke+00}) indicate a hidden AGN in the NGC~7679. Contrary, the NGC~7679 X-ray spectrum is not highly absorbed and $N_\mathrm{H} < 4\times 10^{20}$ cm$^{-2}$ (see discussion in section 4.1). As a matter of fact Bian \& Gu (\cite{BG06}) recently found a very high detectability of hidden BLRs ($\sim 85 $\%) for Compton-thin Sy2s with higher \o3 \ luminosity of $L([\mathrm{O}^{+2}]\lambda5007)\,>\,10^{41}$ erg s$^{-1}$. 

We have to note that NGC~7679 resembles in many respects the galaxy IRAS 12393+3520. In this galaxy direct X-ray evidence suggests the presence of a hidden AGN (Guainazzi et al., \cite{G+00}). This homology can be seen in Fig.~\ref{fig5} where the spectral energy distribution (SED) from the radio to the X-ray band of NGC~7679 is shown. 

The composite nature of NGC 7679 is clearly seen. Whereas the starburst component dominates in the FIR-IR range, the X-ray band emission is well below that of a typical Sy1. The extrapolation of the power-low X-ray spectrum to 13.6 eV shows a much lower value than the typical Sy2 emission at this wavelength. This again favors the idea about a hidden central engine. Guainazzi et al. (\cite{G+00}) suppose that a dusty ionized absorber is able to obscure selectively the optical emission, leaving the X-rays almost unabsorbed.

\subsection{Ionization structure in the circumnuclear region of the NGC~7679} 

The ionization map (the right panel of Fig.~\ref{fig4}) displays the clear signature of highly-excited gas. The \Oc /\ha -ratio increases in the direction of the counterpart galaxy NGC~7682 reaching a maximum of $\approx 2.5$ at about 12 arcsec off the nucleus. More than 15 years ago Durret \& Warin (\cite{DW90}) also reported about the presence of  high-ionization gas in this direction (see their Fig.3a) but their result seemingly did not attract attention. 

On the other hand at PA$\ \approx 0^\circ$ our map shows values around \Oc /\ha $\ \approx 0.3$ and the ionization in this direction is entirely due to the young hot stars.
 
%

   \begin{figure} 
      \centering
      \includegraphics[width=9cm]{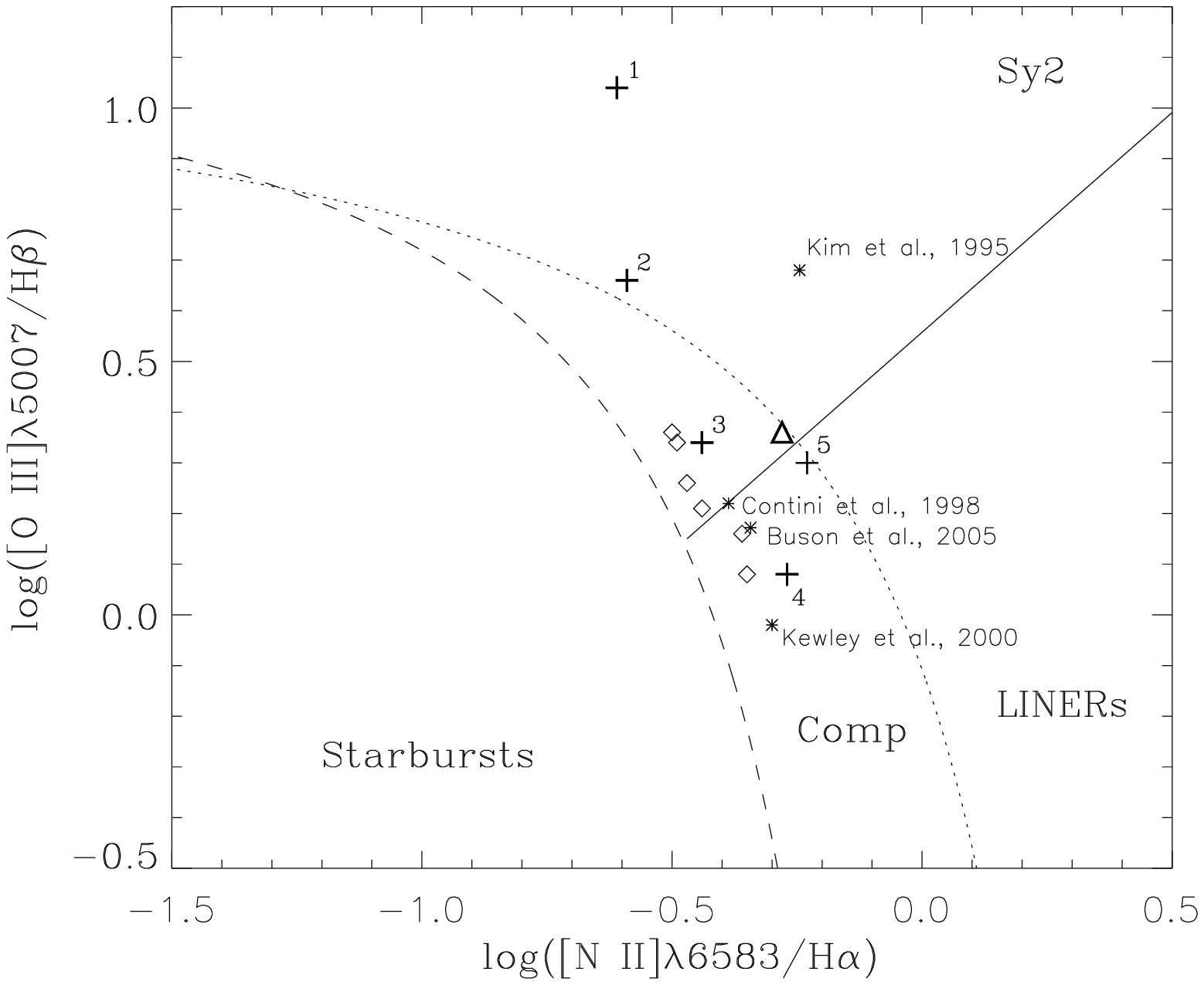}
      \caption {
                The \Oc /\hb \ vs. \Nb /\ha \ diagnostic diagram of Veilleux \& Osterbrock (\cite{VO87}). 
                The dashed and dotted theoretical lines demarcate between Starbursts and AGNs according to 
                Kauffmann et al. (\cite{Kau2003}) and Kewley et al. (\cite{Ke+01}), respectively.
                The line dividing between LINERs and SyGs is taken according to PA02. The label ``Comp''
                indicates the region of the diagram in which composite objects are expected to be found.
                The diagnostic value measured by us is denoted by thick triangle.
                See text for other designations. 
               }
      \label{fig6}
   \end{figure}
%

The \Oc /\hb \ vs.  \Nb /\ha \ diagnostic diagram (Veilleux \& Osterbrock, \cite{VO87}) helps to delineate the different ionization mechanisms maintaining the ionization of gaseous component in AGNs and in Starbursts. In Fig.~\ref{fig6} such a diagram is shown for NGC~7679. Kewley et al. (\cite{Ke+01}) distinguish between Starbursts and AGN using a theoretical upper limit derived from star forming models. This limit is shown as a dotted line in Fig.~\ref{fig6}. Objects with emission-line ratios above this limit cannot be explained by any possible combination of parameters in a star forming model. Kauffmann et al. (\cite{Kau2003}) published an updated estimate for the starburst boundary derived from the SDSS observations. In Fig.~\ref{fig6} this boundary is shown as a dashed line. The location of the Composites is expected to lie between these two lines (see e.g. Panessa et al., \cite{Pa+05}).

In Fig.~\ref{fig6} we plot the emission-line flux ratios of NGC~7679 measured in an aperture of 3 arcsec in steps of 3 arcsec both along the PA$\approx 80^\circ$ (with crosses) and PA$ = 0^\circ$ (with diamonds). The labels 1 - 5 for PA$\approx 80^\circ$ correspond to the labels in the right panel of Fig~\ref{fig4}.  Using spectra taken from the Smithsonian Astrophysical Observatory data Center Z-Machine Archive obtained with 3 arcsec slit width, we estimate the observed $F$(\ha )/$F$(\hb ) $\sim$ 5 in NLR. On Fig.~\ref{fig6} positions 1 and 2 at PA $= 80^\circ$ off the nucleus lie well within the region occupied by the Sy 2 galaxies. The position 5, which is at the same distance from the nucleus but in opposite direction, is located nearly on the dividing line.

All points which refer to the PA $= 0^\circ$ are situated between Kauffmann's and Kewley's demarcation lines in the region of Composites.

In Fig.~\ref{fig6} we also plot with asterisks the nuclear diagnostic ratios according to the data of authors presented in Table~\ref{tab2}. The thick triangle refers to the nucleus according to our measurements under the assumption of $F$(\ha )/$F$(\hb ) = 8.5 (Contini et al., \cite{Cont+98}). The large scattering of nuclear values is probably due to the variations of the strength of \hb \ absorption line of the star-forming stellar population.

\subsection{Unabsorbed SyGs with and without hidden BLRs} 

The unabsorbed Sy2 galaxies with low absorption in X-rays ($N_\mathrm{H}< 10^ {22}$ cm$^{-2}$) possess a hidden or nonhidden central engine and BLRs. We have used the \Oc\, emission to test the presence of hidden or nonhidden AGN sources in  unabsorbed Sy2 galaxies in the sample of PB02 (14 objects) and Panessa et al., \cite{Pa+05} (6 objects selected by Moran et al., \cite{Mo+96}) in the same way as it was done for NGC~7679 (Subsections 4.1 and 4.2). 
We derive the ratio ($N_\mathrm{ph}/N_\mathrm{ion})_{h\nu > 55\,\mathrm{eV}}$ following equations (1) and (2) under the assumtions of $n_\mathrm{e}\approx 5\times 10^4$ cm$^{-3}$ (which is an order of magnitude smaller than the critical electron density for the \Oc \ emission), T$_\mathrm {e}\approx10^4$ K, and CF $\approx 10^{-2}$. These assumptions refer to the inner circumnuclear clouds of AGNs. 

   \begin{table}
      \caption[]{The photon deficiency for unabsorbed Sy2s  
      discussed by Panessa and Bassani (\cite{PB02}) and Panessa et al. (\cite{Pa+05})}
         \label{tab3}
     $$ 
         \begin{tabular}{lll}
            \hline
            \hline \\
galaxy          &  ($N_\mathrm{ph}/N_\mathrm{ion})_{h\nu > 55\,\mathrm{eV}}$ &
                   $N_\mathrm{rec}/N_\mathrm{ion}$ \\
                & & (lower limit) \\   
&&\\ \hline \\
ESO 540-G001             &  4.2 & 13.0 \\ 
CGCG 551-008             &  1.0 & \\      
MCG -03-05-007           &  2.2 & \\      
UGC 03134                & 19.5 & \\      
IRAS 20051-1117          &  1.6 & 1.2\\   
CGCG 303-017             &  1.3 & 2.0\\   
IC 1631          &  0.3 & \\
NGC 2992         &  2.0 & \\
NGC 3147         &  0.4 & \\
NGC 4565         &  6.7 & \\
NGC 4579         &  0.2 & \\
NGC 4594         &  1.7 & \\
NGC 4698         &  0.3 & \\
NGC 5033         &  1.3 & \\
MRK 273x         &  0.4 & \\
NGC 5995         &  0.4 & \\
NGC 6221         &  0.02 & \\
NGC 6251         &  6.0 & \\
NGC 7590         &  0.4 & \\[6pt]
NGC 7679         &  3.4 (2.0 from our data)& \\[6pt]

            \hline
            \hline            
            \noalign{\smallskip}
         \end{tabular}
     $$ 

   \end{table} 
   
The ratios are presented in Table~\ref{tab3}.  For the objects discussed in Panessa et al. (\cite{Pa+05}) the most popular (i.e. as in NED) galaxy names are used. The  $L^\mathrm{corr}([\mathrm{O}^{+2}]\lambda5007)$ values are taken from PB02 and Panessa et al. (\cite{Pa+05}). In the case of NGC~7679 we have used both their and our determinations of  $L^\mathrm{corr}([\mathrm{O}^{+2}]\lambda5007)$. 

For three objects with estimated broad H$\alpha$ component $L^\mathrm{broad}_{\mathrm{H}\alpha}$ (Panessa et al., \cite{Pa+05}, Table 1 therein) we derive also the number of recombinations $N_\mathrm{rec}$ resulting in the \ha \ emission. We assume $T_\mathrm{e}= 10^{4}$ K and CF = 1 which leads to the estimation of the lower limit of the value of $N_\mathrm{rec}$.  The $N_\mathrm{rec}/N_\mathrm{ion}$ lower limits are also presented in Table~\ref{tab3}. 
 
One can see that 17 out of 20 objects of the unabsorbed Sy2s discussed here reveal $N_\mathrm{ph}/N_\mathrm{ion})_{h\nu > 55\,\mathrm{eV}} > 0.3$. This indicates that the central AGN sources in a considerable part of the unabsorbed Sy2s are obscured. The NGC~7679 does not make an exception and also possesses a hidden AGN engine suggested both by the \Oc\ morphology and by the photon deficiency.  

It is still not clear what kind of physical process is related to the presence of hidden central engines in Sy2s.
PB02 suggest two scenarios for the unabsorbed Sy2s (i) the central engine and their BLR must be hidden by an absorbing medium with high value of the $A_\mathrm{V}/N_\mathrm{H}$ ratio, and (ii) the BLR is very weak or absent.
        
\section{Conclusions} 
	
We present a new \Oc \ emission\,-\,line image of the circumnuclear region of NGC~7679 which shows elliptical isophotes extended along the PA\,$\approx 80^\circ \pm 10^\circ$ in the direction to the counterpart galaxy NGC~7682. The maximum of this emission is displaced by about 4 arcsec from the photometric center defined by the continuum emission.	

The ratio of the quantity of ionizing photons inferred from the observed extinction corrected \Oc \ luminosity 
to the number of ionizing photons with $h{\nu}\,> $\,55\,eV provided by the central AGN source  
($N_\mathrm{ph}/N_\mathrm{ion}$)$_{h\nu\,>\,55\,\mathrm{eV}} \approx 0.2 - 20$ as well as the presence of weak and elusive \ha \ broad wings probably indicate a hidden AGN.

The high ionization inferred by the flux ratio \Oc /\ha \ in the direction of about PA$\,\approx\,80^\circ\pm10^\circ$ coincides with the direction to the counterpart galaxy NGC~7682. It is possible that the dust and gas in this direction has a direct view to the central AGN engine. It suggests that starburst and dust decay in this direction have occurred because of tidal interaction between the two galaxies. 

In the direction PA$\approx 0^\circ$ the ionization is entirely caused by hot stars.
  
A large part of the unabsorbed Compton-thin Sy2s with higher \o3 \, luminosity ($\ga 10^{41}$ erg s$^{-1}$) possesses a hidden AGN source.

\begin{acknowledgements}

We are grateful to the referee, Lucio Buson, for his valuable comments which improved both the content and the clarity of this manuscript.

We would like to thank T. Bonev, Institute of Astronomy of Bulgarian Academy of Sciences, for kindly providing the Fabry-Perot observations and for useful discussions. We are grateful to S. Zhekov, Space Research Institute of Bulgarian Academy of Sciences, for the numerous fruitful discussions and especially for the analysis of the X-ray properties of NGC~7679. 
 
Our work was partially based on data from the La Palma ING, ESO NTT, and XMM-Newton Archives.

This research has made use of the SIMBAD database, operated at CDS, Strasbourg, France, and of the NASA/IPAC Extragalactic Database (NED) which is operated by the Jet Propulsion Laboratory, California Institute of Technology, under contract with the National Aeronautics and Space Administration.

We acknowledge the support of the National Science Research Fund by the grant No.F-201/2006.

\end{acknowledgements}



\begin{thebibliography}{}

   \bibitem[1973]{AP73}
      {Aldrovandi, S. M. V., \& Pequignot, D. 1973, A\&A, 25, 137}
   \bibitem[1999]{Ba+99} 
      {Bassani, I., Dadina, M., Maiolino, R.,et al. 1999, ApJS, 121,473}
   \bibitem[2005]{BL05} 
      {Baskin, A. \& Laor, A. 2005, MNRAS, 358, 1043}
   \bibitem[2006]{BG06}
      {Bian, W., \& Gu, Q. 2006, ApJ accepted (astro-ph/0611199)}
   \bibitem[1995]{Bo95}
      {Boyle, B. J., McMahon, R. G., Wilkes, B. J.,\& Elvis, M. 1995, MNRAS, 276, 315}					
   \bibitem[2006]{Bu+06}
      {Buson, L. M., Cappellari, M., Corsini, E. M., Held, E. V., Lim, J., \& Pizzella,  
                   A. 2006, A\&A, 447, 441}
   \bibitem[1991]{Cond+91} 
      {Condon, J., Huang, Z., Yin, Q., \& Thuan, T. 1991, ApJ, 378, 65}
   \bibitem[1998]{Cont+98}       
      {Contini T., Considere S., \& Davoust E. 1998, A\&AS, 130, 285}
   \bibitem[2001]{Ce+01}     
      {Della Ceca, R., Pellegrini, S., Bassani, L., Beckmann, V., Cappi, M., Palumbo, G. 
       G. C., Trinchieri, G., \& Wolter, A. 2001, A\&A, 375, 781 (DC01)}
   \bibitem[1984]{DL+1984}
      {Draine, B. T., \& Lee, H. M. 1984, ApJ, 285, 89}
   \bibitem[1990]{DW90}
      {Durret, F., \& Warin, F. 1990, A\&A, 238, 15}
   \bibitem[1994]{E+94}
      {Elvis M., Wilkes, B. J., McDowell, J. C., Green, R. F., Bechtold, J., Willner, S. P., Oey, M. S., 
      Polomski, E., \& Cutri, R. 1994, ApJS 95, 1}
   \bibitem[1995]{G+95}
      {Golev, V., Yankulova, I., Bonev, T., \& Jockers, K. 1995, MNRAS, 273, 129}
   \bibitem[1996]{G+96}
      {Golev, V., Yankulova, I., \& Bonev, T. 1996, MNRAS, 280, 29}
   \bibitem[1994]{GD94}
      {Granato, G. L., \& Danese, L. 1994, MNRAS, 268, 235} 
   \bibitem[1996]{Gr+96}
      {Griffiths, R. E., Della Ceca, R., Georgantopoulos, I., Boyle, B., Stewart, G.,
       Shnks, T., \& Fruscione, A. 1996 MNRAS, 281, 71} 
   \bibitem[2006]{Gu+06}
      {Gu, Q., Melnick, J., Fernandes, R. Cid, Kunth, D., Terlevich, E., \& Terlevich, R.
      2006, MNRAS, 366, 480}
   \bibitem[2001]{Gu+01}
      {Gu, Q. S., Huang, J. H., de Diego, J. A., Dultzin-Hacyan, D., Lei, S. J., \& 
       Benitez, E. 2001, A\&A, 374, 932}
   \bibitem[2000]{G+00}
      {Guainazzi, M., Dennefeld, M., Piro, L., Boller, T., Rafanelli, P., \& Yamauchi, M. 
      2000, A\&A, 355, 113}
   \bibitem[1990]{He+90}
      {Heckman, T. M., Armus, L., \& Miley, G. K. 1990, ApJS, 74, 833}
   \bibitem[1997]{Jo97} 
      {Jockers, K. 1997, Experimental Astronomy, 7, 305}
   \bibitem[2000]{Jo+00} 
      {Jockers, K., Credner, T., Bonev, T., Kiselev, N., Korsun, P., Kulik, I., Rosenbush, V., Andrienko, A., 
       Karpov, N., Sergeev, A., \& Tarady, V. 2000, Kinematika i Fizika Nebesnykh Tel, Suppl, No. 3, 13}
   \bibitem[2003]{Kau2003}
      {Kauffmann, G., Heckman, T. M., Tremonti, C., et al. 2003, MNRAS, 346, 1055}
   \bibitem[2000]{Ke+00}
      {Kewley, L. J., Heisler, C. A., Dopita, M. A., Sutherland, R.
      Norris, R., Reynolds, J., \& Lumsden, S. 2000, ApJ, 530, 704}
   \bibitem[2001]{Ke+01}      
      {Kewley, L. J., Heisler, C. A., Dopita, M. A., \& Lumsden, S. 2001, ApJS, 132,           37}
   \bibitem[1995]{Kim+95}
      {Kim, D.-C., Sanders, D. B., Veilleux, S., Mazzarella, J. M., \& Soifer, B. T.     
           1995, ApJS, 98, 129}   
   \bibitem[1995]{KP95}    
      {Kotilainen, J. K., \& Prieto, M. A. 1995, A\&A, 295, 646}
   \bibitem[2001]{Le+01}
      {Levenson, N., Weaver, K., \& Heckman, T. 2001, ApJ, 550, 230}
   \bibitem[1991]{Li+91}
      {Lipari, S., Bonatto, Ch., \& Pastoriza, M. 1991, MNRAS, 253, 19}
   \bibitem[1995]{M-H+95}      
      {Mas-Hesse, J. M., Rodriguez-Pascual, P. M., Sanz Fernandez de Cordoba, L., Mirabel, I. F., 
      Wamsteker, W., Makino, F., \& Otani, C. 1995, A\&A 298, 22}
   \bibitem[1996]{Mo+96}
      {Moran, E. C., Halpern, J. P.,\& Helfand, D. J. 1996, ApJS, 106, 341}
   \bibitem[2006]{MK06}
      {Moustakas, J., \& Kennicutt, R. C. 2006, ApJS, 164, 81}								
   \bibitem[1989]{Os89}
      {Osterbrock, D. 1989, Astrophysics of gaseous nebulae and active galactic nuclei,  
                       University Science Books}   
    \bibitem[2002]{PB02}
      {Panessa, F., \& Bassani, L. 2002,  A\&A, 394, 435 (PB02)}
   \bibitem[2005]{Pa+05}
      {Panessa, F., Wolter, A., Pellegrini, S., Fruscione, A., Bassani, L., Della Ceca, R., 
       Palumbo, G., \& Trinchieri, G. 2005, ApJ, 631, 707}
   \bibitem[1992]{PK92}
      {Pier, E. A., \& Krolik, J. 1992, ApJ, 401, 99} 
   \bibitem[1989]{Po89}
      {Pogge, R. W. 1989, AJ, 98, 124}
   \bibitem[1999]{Ri+99}
      {Risaliti, G., Maiolino, R., \& Salvati,  M. 1999, ApJ, 522, 157} 
   \bibitem[2002]{Ri02}
      {Risaliti, G. 2002,  A\&A, 386, 379}    
   \bibitem[2001]{Ros+01} 									
   		{Rosati, P., \& Chandra Deep Field South Team, 2001, A\&AS, Bull.AAS, 33, 1519}                     
   \bibitem[1988]{San+88}
      {Sanders, D., Soifer, B., Elias, J., Madore, B., Matthews, K., Neugebauer, G., \&           
       Scoville, N. 1988, ApJ, 325, 74}
   \bibitem[1997]{Sch+97}
      {Schmitt, H. R., Kinney, A. L., Calzetti, D., \& Storchi Bergmann, T.  1997, AJ 114, 592}
   \bibitem[1996]{Sim+96}
      {Simpson, C., Mulchaey, J. S., Wislon, A. S., Ward, M. J., \& Alonso-Herrero, A. 1996, ApJ, 457, L19}
   \bibitem[1997]{Sim+97}
      {Simpson, C., Wislon, A. S., Bower, G., Heckman, T. M., Krolik, J. H., \& Miley, G. K. 
      1997, ApJ, 474, 121}
   \bibitem[1998]{Sm+98}
      {Smith, H. E., Lonsdale, C. J., \& Londsdale C. J. 1998, ApJ, 492, 137} 
   \bibitem[1992]{St92}
      {Stine, P. C. 1992, ApJS, 81, 49} 
   \bibitem[1995]{T+93}   
      {Telesco, C. M., Dressel, L., \& Wolstencroft,R. 1993, ApJ, 414, 120}
   \bibitem[1995]{Ve+95}
      {Veilleux, S., Kim, D.-C., Sanders, D. B., Mazzarella, J. M., \& Soifer, B. T. 1995, ApJS, 98, 171}
   \bibitem[1987]{VO87} 
      {Veilleux, S., \& Osterbrock, D.\,E. 1987, ApJS, 63, 295}
   \bibitem[1993]{Wi+93}
      {Wilson, A. S., Braatz, J. A., Heckman, T. M., Krolik, J. H., \& Miley, G. K. 1993, ApJ, 419, L61} 
   \bibitem[1999]{Yan99}
      {Yankulova, I. 1999, A\&A, 344, 36}    
                 
\end{thebibliography}
\end{document}